\documentclass[%
 reprint,
amsmath,amssymb,
 aps,
]{revtex4-2}
\usepackage{ytableau}
\usepackage[hyperfootnotes=false]{hyperref}
\hypersetup{
 colorlinks=true,
 citecolor=blue,
 linkcolor=red,
 urlcolor=blue}
\usepackage{mathtools}
\usepackage{nccmath} 
\usepackage{xcolor}
\usepackage{caption}
\usepackage{subcaption}
\usepackage{graphicx,physics}% Include figure files
\usepackage{dcolumn}% Align table columns on decimal point
\usepackage{bm}% bold math
\begin{document}

\preprint{APS/123-QED}

\title{Masses and Magnetic Moments of Singly Heavy Pentaquarks using the Gursey-Radicati Mass Formula, Effective Mass, and Screened Charge Scheme}% Force line breaks with \\
\author{Ankush Sharma, Alka Upadhyay}
 \affiliation{Department of Physics and Material Sciences, Thapar Institute of Engineering and Technology, Patiala, India, 147004}
 
\email{ankushsharma2540.as@gmail.com}

\date{\today}% It is always \today, today,
\begin{abstract}
Motivated by the recent discovery of single heavy tetraquark structures, $T_{c\bar{s}0}^a (2900)^{++}$ and $T_{c\bar{s}0}^a(2900)^0$ by the LHCb collaboration, masses and magnetic moments of singly heavy pentaquark states are estimated in this work. To classify the singly heavy pentaquark structures, we employ the special unitary representation. By using the SU(3) flavor representation, singly heavy pentaquark states are classified into the allowed flavor multiplets. Also, by using the extension of the Gursey-Radicati mass formula and the effective mass scheme, masses of singly heavy pentaquark states are estimated. Further, magnetic moments of singly heavy pentaquarks have been calculated using the effective mass and the screened charge schemes. A thorough comparison of our results shows reasonable agreement with the available theoretical data and may be helpful for future experimental studies.
\end{abstract}

\maketitle

%\tableofcontents

\section{Introduction}
 In recent times, the study of exotic hadrons has been developed widely. Many experimental observations about tetraquarks and pentaquarks have been reported by different facilities like LHCb, Belle, ATLAS, and CMS collaborations. In this work, we investigate the masses and magnetic moments of singly heavy pentaquark states. Pentaquarks are exotic hadrons made up of five quarks. They were predicted by theory in the 1960s, but their existence was not confirmed experimentally until 2002. Pentaquarks are an important area of research in particle physics, as their study can help us understand the fundamental forces of nature and the structure of matter. In 2004, The H1 Collaboration at
HERA has reported a narrow resonance $\Theta_c$ appearing
in $D^{*-}p$ with quark content $uudd\Bar{c}$ and $D^{*+}\Bar{p}$ with quark content $\Bar{u}\Bar{u}\Bar{d}\Bar{d}c$ states produced in inelastic $ep$ collisions, with a mass of
$3099\pm 3 \pm 5$ MeV and width of $12 \pm 3$ MeV \cite{Experimental}. Various models have also supported that instead of $\Bar{s}$ quark in the $\Theta^+$ pentaquark, a $\Bar{c}$ or $\Bar{b}$ is more likely to be bound \cite{Support1, Support2}. This is more supported by different theoretical models that singly heavy tetraquarks are more likely to be found \cite{theory1,theory2}. Also, recent observations of singly heavy tetraquark states by LHCb collaboration serve as the motivation and open the window for upcoming observations of singly heavy pentaquark states. Recently, in 2022, LHCb collaboration observed two singly heavy tetraquark states having quark content $u\bar{d}c\bar{s}$ and $\bar{u}dc\bar{s}$ with the significance of 
6.5 $\sigma$ and 8 $\sigma$ respectively \cite{2900}. These two tetraquark candidates $T_{c\bar{s}0}^a (2900)^{++}$ and $T_{c\bar{s}0}^a(2900)^0$ were observed in the $D_s^+\pi^+$ and $D_s^+\pi^-$ invariant mass spectra in two $B$-decay processes $B^+ \rightarrow D^-D_s^+\pi^+$ and $B^0 \rightarrow \bar{D^0}D_s^+ \pi^-$, respectively. The isospin and spin-parity quantum numbers are determined to be $(I)J^P = (1)0^+$. These two states should correspond to two different charged states of the isospin triplet. The measured mass and width are:
\begin{align}
   M_{exp} = 2908\pm 11 \pm 20 MeV, \hspace{0.3cm}
   \Gamma_{exp}= 136 \pm 23 \pm 13  MeV.  
\end{align}
Also, a hidden-charm pentaquark structure $P_{\psi s}^\Lambda(4338)^0$ having minimal quark content $udsc\bar{c}$, is observed in the $J/\psi\Lambda$ mass in the $B^-\rightarrow J/\psi\Lambda p$ decays observed with a statistical significance of 15 $\sigma$, far beyond the required five standard deviations to claim the observation of a particle \cite{4338}. The mass and the width of the new pentaquark are measured to be $4338.2\pm0.7\pm0.4$ MeV and $7.0\pm1.2\pm1.3$ MeV, respectively. The preferred quantum numbers are $J^P = 1/2^-$. Similarly, many other hidden-charm pentaquark structures such as $P_{cs}(4459)$, $P_c(4312)^+$, $P_c(4380)$ and $P_c(4450)$ are discovered in the recent decade \cite{20211278, PhysRevLett.122.222001, PhysRevLett.115.072001}. These discoveries of exotic hadrons inspired theorists for the spectroscopy of these states. By taking the motivation from the discovery of singly heavy tetraquark structures, we estimated the masses and magnetic moments of singly heavy pentaquarks. The field of singly heavy pentaquarks has grown significantly and many works have been proposed to study their dynamics \cite{PRD, PRD2, COhen,2004a}. In Ref. \cite{2004a}, a study of singly heavy pentaquarks in anti-sextet configuration is done using the various clustered quark models. Using the extension of the Gursey-Radicati-inspired mass formula, we calculated the masses of singly heavy pentaquark states. The extension of the Gursey-Radicati mass formula seems a very useful tool to study the hadron masses \cite{FG}. Many works have been proposed using the extension of the GR mass formula for the successful prediction of exotic hadron masses \cite{Santo, HOLMA, Sharma_2023, Ankush}. Further, pentaquarks are exotic hadrons consisting of four quarks and an antiquark, exhibit a magnetic moment that holds special significance owing to their unique composition. Exploring the magnetic moments of hadrons provides valuable insights into their internal structure and the behaviors of their constituent quarks. By serving as sensitive probes for charge and current distribution within these particles, magnetic moments contribute to testing Quantum Chromodynamics (QCD) predictions. Additionally, they offer a means to unveil information about the breaking of symmetries within strong forces. The study of magnetic moment is essential for understanding how pentaquarks respond to magnetic influences and contributes significantly to the broader understanding of their characteristics. In essence, the magnetic moment becomes a measure of the strength of a pentaquark's magnetic field, dictated by the interplay between the spin and orbit of the constituent quarks forming the particle. Therefore, we estimated the magnetic moments for singly heavy pentaquark states using the effective mass and screened charge scheme.\\
This work is organized as follows: Section II briefly describes the theoretical formalism which includes the classification scheme for singly heavy pentaquarks using the SU(3) flavor representation, an extension of the Gursey-Radicati mass formula, the effective mass scheme, and the screened charge scheme. Section III describes the results discussion and conclusion.

\section{Theoretical Formalism}

\subsection{Classification Scheme for Pentaquarks}
To classify the pentaquark states with a single heavy quark, we utilize the SU(3) flavor representation. The SU(3) flavor representation is the mathematical framework that describes the flavor quantum numbers of quarks in the context of the strong nuclear force, which is mediated by the exchange of gluons. SU(3) is a mathematical symmetry group that plays a crucial role in the description of quark flavors. Pentaquarks with a single heavy quark are particularly interesting because they can provide insights into the behavior of heavy quarks within hadrons. The presence of four light quarks ($u$, $d$, $s$) gives us the following multiplets by assigning fundamental (3) representation to the quarks and ($\Bar{3}$) to the anti-quarks in case of singly heavy pentaquark ($qqqq\Bar{Q}$):
\begin{align}
    [3]\otimes [3]\otimes [3]\otimes [3] =  3 [3] \oplus 2 [\Bar{6}] \oplus 3 [15^{'}] \oplus[15]
    \label{2}
\end{align}
Similarly, for three light quarks and a light anti-quark inside a singly heavy pentaquarks ($qqq\Bar{q}Q$), we get:
\begin{align}
    [3]\otimes [3]\otimes [3]\otimes [\Bar{3}] =  3 [\Bar{3}] \oplus 3 [6] \oplus 2 [15^{''}] \oplus [24]
    \label{3}
\end{align}
By using Eq. \eqref{2} and \eqref{3}, we classified the singly heavy pentaquarks with configurations $qqqq\Bar{Q}$ into the $15^{'}$-plet and $qqq\Bar{q}Q$ into $15^{''}$ -plet representation. $15^{'}$ representation is a mixed symmetric state corresponding to a configuration with symmetry among the particles in the top row, while the particle in the second row introduces a differentiation in the Young tableau representation. $15^{''}$ represents another mixed symmetry, but here, we see a two-row configuration with more complexity: three boxes in the first row and two in the second. This might correspond to states where two particles share one symmetry property and three others share a different one.  Further, we used the SU(2) spin representation and the Young Tableau technique to define the spin wave function of pentaquarks. In SU(2) spin representation, each quark and an antiquark is assigned by a fundamental (2) representation, therefore:
\begin{align}
    [2] \otimes [2] \otimes [2] \otimes [2] \otimes
[2] =  [6]  \oplus 4[4] \oplus 5[2]
\end{align}
the configurations [6], [4], and [2] correspond to the spin values $s$ = 5/2, 3/2, and 1/2, respectively. In spin space, by using the Young tableau technique for the pentaquark system, the spin wavefunction is reported in Table I \cite{young}:
\begin{table}[h]
\centering
    \caption{Young Tableau representation for Pentaquarks \cite{young}.}
    \begin{tabular}{ccc}
     \begin{ytableau}
     \none &
\end{ytableau} $\otimes$
\begin{ytableau}
     \none &
\end{ytableau} $\otimes$
\begin{ytableau}
     \none &
\end{ytableau}  $\otimes$
 \begin{ytableau}
     \none &
\end{ytableau} $\otimes$
\begin{ytableau}
     \none &
\end{ytableau} = \nonumber \\ 
\\
\begin{ytableau}
    \none & & & & &
\end{ytableau} $\oplus$ 
4\begin{ytableau}
    \none & & & & \\
    \none &
\end{ytableau} $\oplus$
5\begin{ytableau}
    \none & & & \\
    \none & &
\end{ytableau}\\ 
    \end{tabular}
\end{table}
For the spin-5/2 case, we can write the spin wavefunction as:
\begin{center}
  \begin{ytableau}
    \none & 1 & 2 & 3 & 4 & 5
\end{ytableau}${\chi_1}^P$ 
\end{center}
similarly, for spin-3/2 pentaquarks, spin wavefunction in terms of Young tableau representation is shown in Table \ref{32}.
\begin{table}[h!]
    \centering
    \caption{Young Tableau representation for spin-3/2 Pentaquarks \cite{young}.}
    \begin{tabular}{cc}
     \begin{ytableau}
    \none & 1 & 2 & 3 & 4 \\
    \none & 5
\end{ytableau}$\chi_2^P$ & \begin{ytableau}
    \none & 1 & 2 & 3 & 5 \\
    \none & 4
\end{ytableau}$\chi_3^P$  \\ &\\
     \begin{ytableau}
    \none & 1 & 2 & 4 & 5 \\
    \none & 3
\end{ytableau}$\chi_4^P$ &
  \begin{ytableau}
    \none & 1 & 3 & 4 & 5 \\
    \none & 2
\end{ytableau}$\chi_5^P$
\end{tabular}
\label{32}
\end{table}

\begin{table}[h]
    \centering
    \caption{Young Tableau representation for spin-1/2 Pentaquarks \cite{young}.}
    \begin{tabular}{ccc}
       \begin{ytableau}
    \none & 1 & 2 & 3  \\
    \none & 4 & 5
\end{ytableau}$\chi_6^P$ 
 
  \begin{ytableau}
    \none & 1 & 2 & 4  \\
    \none & 3 & 5
\end{ytableau}$\chi_7^P$\\ 
\\
\hspace{0.3cm}
  \begin{ytableau}
    \none & 1 & 3 & 4 \\
    \none & 2 & 5
\end{ytableau}$\chi_8^P$ 
  \begin{ytableau}
    \none & 1 & 2 & 5  \\
    \none & 3 & 4
\end{ytableau}$\chi_9^P$ 
  \begin{ytableau}
    \none & 1 & 3 & 5  \\
    \none & 2 & 4
\end{ytableau}$\chi_{10}^P$ 
    \end{tabular}
    \label{12}
\end{table}
Further, for the spin-1/2 case, the spin wave function is represented in Table \ref{12} by using the Young tableau techniques. Ten symmetries are available from $\chi_1$ to $\chi_{10}$ for spin 5/2 to spin 1/2, respectively. They have their associated symmetries as follows: \\
$\chi_1$  is symmetric w.r.t five quarks. $\chi_2$ is symmetric with respect to the first four quarks and mixed symmetry with respect to the fifth quark. $\chi_3$ is symmetric among the first three quarks, with mixed symmetry with respect to the rest. $\chi_4$ have mixed symmetry overall. $\chi_5$ is anti-symmetric under the exchange of the first two quarks. $\chi_6$ have mixed symmetry due to a complex combination of terms. $\chi_7$ is predominantly symmetric. $\chi_8$ is anti-symmetric, as indicated by structure and sign changes upon exchange. $\chi_9$ have mixed symmetry, combining symmetric and antisymmetric components. $\chi_{10}$ is anti-symmetric, clear from term structure and sign changes. Using the $\chi_4$ symmetry, we calculated the magnetic moments of spin-3/2 pentaquarks, and a linear combination of $\chi_6$ and $\chi_9$ symmetries is used for the magnetic moments of spin-1/2  pentaquarks, with the help of effective mass and screened charge schemes, described in subsections C and D. To calculate the mass spectrum of the pentaquark states, we used the extension of the Gursey-Radicati mass formula which is discussed in the next subsection. The corresponding spin wavefunctions for these ten spin symmetries are written as \cite{magneticmom}: 
\begin{widetext}
\begin{align}
     \chi_1 =& \ket{\uparrow\uparrow\uparrow\uparrow\uparrow} \\
     \chi_2 =& \frac{2}{\sqrt{5}}\ket{\uparrow\uparrow\uparrow\uparrow\downarrow - \frac{\sqrt{5}}{10}(\uparrow\uparrow\uparrow\downarrow\uparrow + \uparrow\uparrow\downarrow\uparrow\uparrow + \uparrow\downarrow\uparrow\uparrow\uparrow + \downarrow\uparrow\uparrow\uparrow\uparrow)}\\
    \chi_3 =& \frac{\sqrt{3}}{2}\ket{\uparrow\uparrow\uparrow\downarrow\uparrow - \frac{1}{2\sqrt{3}}(\uparrow\uparrow\downarrow\uparrow\uparrow + \uparrow\downarrow\uparrow\uparrow\uparrow + \downarrow\uparrow\uparrow\uparrow\uparrow)}\\
   \chi_4 =& \frac{1}{\sqrt{6}}\ket{(2\uparrow\uparrow\downarrow\uparrow\uparrow - \uparrow\downarrow\uparrow\uparrow\uparrow - \downarrow\uparrow\uparrow\uparrow\uparrow)}\\ 
   \chi_5 =& \frac{1}{\sqrt{2}}\ket{(\uparrow\downarrow\uparrow\uparrow\uparrow - \downarrow\uparrow\uparrow\uparrow\uparrow)}\\
   \chi_6 =& \frac{1}{3\sqrt{2}}\ket{(\uparrow\downarrow\downarrow\uparrow\uparrow + \downarrow\uparrow\downarrow\uparrow\uparrow + \downarrow\downarrow\uparrow\uparrow\uparrow - \uparrow\uparrow\downarrow\downarrow\uparrow - \uparrow\downarrow\uparrow\downarrow\uparrow 
   - \downarrow\uparrow\uparrow\downarrow\uparrow - \uparrow\uparrow\downarrow\uparrow\downarrow - \uparrow\downarrow\uparrow\uparrow\downarrow - \downarrow\uparrow\uparrow\uparrow\downarrow) + \frac{1}{\sqrt{2}} \uparrow\uparrow\uparrow\downarrow\downarrow}\\
   \chi_7 =& \frac{1}{3} \ket{(2\uparrow\uparrow\downarrow\uparrow\downarrow - \uparrow\downarrow\uparrow\uparrow\downarrow - \downarrow\uparrow\uparrow\uparrow\downarrow - \uparrow\uparrow\downarrow\downarrow\uparrow + \downarrow\downarrow\uparrow\uparrow\uparrow) + \frac{1}{6}(\uparrow\downarrow\uparrow\downarrow\uparrow - \uparrow\downarrow\downarrow\uparrow\uparrow - \downarrow\uparrow\downarrow\uparrow\uparrow + \downarrow\uparrow\uparrow\downarrow\uparrow)} \\
   \chi_8 =& \frac{1}{\sqrt{3}} \ket{(\uparrow\downarrow\uparrow\uparrow\downarrow - \downarrow\uparrow\uparrow\uparrow\downarrow) -\frac{1}{2\sqrt{3}}(\uparrow\downarrow\uparrow\downarrow\uparrow + \uparrow\downarrow\downarrow\uparrow\uparrow - \downarrow\uparrow\downarrow\uparrow\uparrow - \downarrow\uparrow\uparrow\downarrow\uparrow)}\\
 \chi_9 =& \frac{1}{\sqrt{3}}\ket{(\uparrow\uparrow\downarrow\downarrow\uparrow + \downarrow\downarrow\uparrow\uparrow\uparrow\uparrow) -\frac{1}{2\sqrt{3}}(\uparrow\downarrow\uparrow\downarrow\uparrow + \uparrow\downarrow\downarrow\uparrow\uparrow + \downarrow\uparrow\downarrow\uparrow\uparrow + \downarrow\uparrow\uparrow\downarrow\uparrow )}\\
 \chi_{10} =& \frac{1}{2}\ket{(\uparrow\downarrow\uparrow\downarrow\uparrow - \uparrow\downarrow\downarrow\uparrow\uparrow + \downarrow\uparrow\downarrow\uparrow\uparrow - \downarrow\uparrow\uparrow\downarrow\uparrow)}
\end{align}
\end{widetext}
Furthermore, it should be noted that the color wave function associated with a bound state of a hadron is subject to color confinement, resulting in color singlet states. The color wavefunction in terms of the Young Tableau scheme is represented in Ref.\cite{magneticmom, colorwavefunction}. Furthermore, we focus on ground-state pentaquark states ($s$-wave). Consequently, the symmetrical restriction for the spatial wave function is trivial. We can obtain the expressions for magnetic moments reported in Table \ref{tab: expressions} using these spin-flavor wavefunctions. With the help of these symmetries, we obtained the magnetic moment assignments. 
\subsection{The Extended Gursey-Radicati mass formula}
To study the mass spectra of exotic states, the extension of the GR mass formula is a useful method. It was first introduced by F. Gursey and L. Radicati to study the masses of baryons \cite{FG}. Further, it was extended by E. Santopinto and A. Giachino to study the masses of hidden-charm pentaquark states, and the simplest form for the  mass formula is \cite{Santo}:
\begin{align}
  M_{GR} = M_0 &+ AS(S+1) + DY  + E[I(I+1) -1/4 Y^2] \nonumber \\ &+ G C_2(SU(3)) + F N_c
\end{align}
This form of the mass formula can successfully predict the mass spectrum of hadrons containing the charm quarks. This formula was more generalized by P. Holma and T. Ohlson to include the contribution of the bottom quark by generalizing the counter term \cite{HOLMA}:
 \begin{align}
  M_{GR} = \xi M_0 &+ AS(S+1) + DY  + E[I(I+1)  -1/4 Y^2] \nonumber \\ &+ G C_2(SU(3)) + \sum_{i=c,b} F_i N_i
  \label{mass formula}
  \end{align}
Here, $M_0$ is the scale parameter, and $\xi$ is the correction factor to the scale parameter, related to the number of quarks making up the hadron. As each quark contributes 1/3 to the mass of the hadron, therefore, $\xi$ has a value of 1 for baryons ($qqq$), 4/3 for tetraquarks ($q\Bar{q}q\Bar{q}$), and 5/3 for the pentaquarks ($qqqq\Bar{q}$). The spin, isospin, and hypercharge quantum numbers are denoted as $S$, $I$, and $Y$, respectively. $N_i$ stands for the number of charm (anti-charm) $c$($\overline{c}$) or bottom (anti-bottom) $b$($\Bar{b}$) quarks (anti-quarks). Gursey-Radicati mass formula provides a relation between the mass of a particle and its associated quantum numbers (spin, isospin, hypercharge, etc.), offering insights into the interplay between these fundamental properties. Mass formula parameters are taken from Ref. \cite{Santo, HOLMA} where they evaluated the mass formula parameters by using the known baryon spectrum because of less availability of the experimentally available pentaquark states. Mass formula parameters are listed in Table \ref{tab:2}.

\begin{table}[ht]
\centering
\caption{Values of parameters used in the extended GR mass formula with corresponding uncertainties. \cite{Santo}}
\tabcolsep 0.4mm  
\begin{tabular}{cccccccc}
       \hline
       \hline
         & $M_0$ & A & D & E & G & $F_c$ & $F_b$ \\
         \hline
        Values[MeV] & 940.0 & 23.0 & -158.3 & 32.0 & 52.5 & 1354.6  & 4820 \cite{HOLMA} \\
        \hline
        Uncert.[MeV] & 1.5 & 1.2 & 1.3 & 1.3 & 1.3 & 18.2  & 34.4  \\
        \hline
        \hline
       \end{tabular}
        \label{tab:2}
   \end{table}

By using eq. \eqref{mass formula}, the mass spectrum of singly charm and singly bottom pentaquarks are calculated and mentioned in Tables \ref{tab:nc} and \ref{tab:nb}, respectively. The effective mass and screened charge scheme are introduced in the next subsection to calculate the magnetic moment of the singly charm and bottom pentaquark states. 

\begin{table*}\renewcommand{\arraystretch}{0.8}
 \tabcolsep 0.5mm      
       \centering
       \caption{Pentaquark masses for $N_c$ = 1  for all possible configurations of singly charm pentaquarks. Masses are in the units of MeV.}
       \begin{tabular}{ccccccccc}
\hline
\hline
 Quark Content &  S &  I &  Y & $SU(3)_f$ & $C_2(SU(3)_f)$ & $N_c$ & Our prediction(MeV) &  Reference \\
  \hline
  \\
   \\
$P_{c0}^{++}$,  $P_{c0}^{+}$,  $P_{c0}^{0}$,  $P_{c0}^{-}$,  $P_{c0}^{--}$ & 1/2 & 2 & 4/3 & $[4]_{15}$ &  16/3 & 1 & 3185.64 $\pm$ 20.91 & 3487 \cite{Singly}\\
  
   & 3/2 & 2 & 4/3 & $[4]_{15}$ &  16/3 & 1 & 3254.64 $\pm$ 21.37 & 3381 \cite{Singly} \\
  
  \\
  \hline
  \\
  $P_{c1}^+$, $P_{c1}^0$, $P_{c1}^-$, $P_{c1}^{--}$ & 1/2 & 3/2 & 1/3 & $[4]_{15}$ &  16/3 & 1 & 3285.22 $\pm$ 20.14 & 3246 \cite{Singly}\\
  
  & 3/2 & 3/2 & 1/3  & $[4]_{15}$ &  16/3 & 1 & 3354.22 $\pm$ 20.62 & 3343 \cite{Singly}\\
 
 \\
  \hline
  \\
 $P_{c2}^{0}$,  $P_{c2}^{-}$,  $P_{c2}^{--}$ & 1/2 & 1 & -2/3 & $[4]_{15}$ &  16/3 & 1 & 3383.32 $\pm$ 19.72 & 3350 \cite{Singly}\\
  & 3/2 & 1 & -2/3 & $[4]_{15}$ &  16/3 & 1 & 3452.32 $\pm$ 20.12 & 3505 \cite{Singly}\\
 
  \\
  \hline
  \\
  $P_{c3}^{-}$,  $P_{c3}^{--}$ & 1/2 & 1/2 & -5/3 & $[4]_{15}$ &  16/3 & 1 & 3483.06 $\pm$ 19.67 & 3465 \cite{Singly}\\
  & 3/2 & 1/2 & -5/3 & $[4]_{15}$ &  16/3 & 1 & 3552.06 $\pm$ 20.15 & 3655 \cite{Singly}\\
  
  \\
  \hline
  \\
  $P_{c4}^{--}$ & 1/2 & 0 & -8/3 & $[4]_{15}$ &  16/3 & 1 & 3582.80 $\pm$ 19.98 & 3950 \cite{Singly}\\
  & 3/2 & 0 & -8/3 & $[4]_{15}$ &  16/3 & 1 & 3651.80 $\pm$ 20.46 & 3842 \cite{Singly}\\
    \\
 \hline
 \\
 $uuu\bar{u}c$, $uud\bar{u}c$, $udd\bar{u}c$ &   1/2 &  0 &  2/3 & $[4]_{15}$ & 16/3 & 1&  3109.42$\pm$ 19.57 \\
  $ddd\bar{u}c$, $uud\bar{d}c$  &   1/2 &  1 &  2/3 & $[4]_{15}$ & 16/3 & 1&  3173.42$\pm$ 19.72 \\
  $udd\bar{d}c$, $uuu\bar{d}c$  &   1/2 & 2 &  2/3 & $[4]_{15}$ & 16/3 & 1&  3301.42$\pm$ 21.01 \\
 $uus\bar{s}c$, $uds\bar{s}c$, $dds\bar{s}c$ &  3/2 &  0 &   2/3 & $[4]_{15}$ & 16/3 & 1  & 3178.42$\pm$ 20.06  \\
  &  3/2 &  1 &   2/3 & $[4]_{15}$ & 16/3 & 1  & 3242.42$\pm$ 20.21  \\
  &  3/2 &  2 &   2/3 & $[4]_{15}$ & 16/3 & 1  & 3370.42$\pm$ 21.47 \\
 \hline
 \\
  $uuu\bar{s}c$, $uud\bar{s}c$ &  1/2 &  1/2 &  5/3 & $[4]_{15}$ & 16/3 & 1  & 2956.45$\pm$ 19.67 \\
 $udd\bar{s}c$, $ddd\bar{s}c$ &  1/2 &  3/2 &  5/3 & $[4]_{15}$ & 16/3 & 1  & 3052.45$\pm$ 20.07\\
  &  1/2 &  5/2 &  5/3 & $[4]_{15}$ & 16/3 & 1  & 3212.45$\pm$ 22.28 \\
  &  3/2 &   1/2 &   5/3 & $[4]_{15}$ & 16/3 &  1  & 3035.45$\pm$ 20.16 \\
  &  3/2 &   3/2 &   5/3 & $[4]_{15}$ & 16/3 &  1  & 3121.45$\pm$ 20.54 \\
  &  3/2 &   5/2 &   5/3 & $[4]_{15}$ & 16/3 &  1  & 3281.45$\pm$ 22.71 \\
    \\
    \hline 
    \\
 $uus\bar{u}c$, $uss\bar{s}c$, $dss\bar{s}c$ &  1/2 &  1/2 &  -1/3 & $[4]_{15}$ & 16/3 & 1  & 3294.39$\pm$ 19.58 \\
 $dds\bar{u}c$, $dds\bar{d}c$ &  1/2 &  3/2 &   -1/3 & $[4]_{15}$ & 16/3 & 1  & 3390.39$\pm$ 20.14 \\
 $uus\bar{d}c$, $uds\bar{u}c$ &  1/2 &  5/2 &   -1/3 & $[4]_{15}$ & 16/3 & 1  & 3550.39$\pm$ 22.60 \\
  & 3/2 &  1/2 &   -1/3 & $[4]_{15}$ & 16/3 & 1  & 3363.39$\pm$ 20.07  \\
  &  3/2 &  3/2 &   -1/3 & $[4]_{15}$ & 16/3 & 1  & 3459.39$\pm$ 20.62  \\
  & 3/2 &  5/2 &   -1/3 & $[4]_{15}$ & 16/3 & 1  & 3619.39$\pm$ 23.03  \\
    \\
  \hline
  \\
 $uss\bar{u}c$, $uss\bar{d}c$ &  1/2  &  0 &  -4/3 & $[4]_{15}$ &  16/3 &  1  & 3415.35$\pm$ 19.63 \\
  $dss\bar{d}c$ &  1/2  &  1 &  -4/3 & $[4]_{15}$ &  16/3 &  1  & 3479.35$\pm$ 19.73 \\
  &   3/2  & 0 & -4/3 & $[4]_{15}$ &   16/3 & 1  & 3484.35$\pm$ 20.12\\
  &   3/2  & 1 &  -4/3 & $[4]_{15}$ &   16/3 & 1  & 3548.35$\pm$ 20.22\\
   \\
  \hline
  \\
 $sss\bar{u}c$, $sss\bar{d}c$ &   1/2  &  1/2 &  -7/3 & $[4]_{15}$ & 16/3 &   1  & 3568.32$\pm$ 19.80\\
 &   3/2  &  3/2 &   -7/3 & $[4]_{15}$ & 16/3 &  1  &  3637.32$\pm$ 20.28\\
 
  \\
  \hline
  \hline
 \end{tabular}
   \label{tab:nc}
\end{table*}

 \begin{table*}\renewcommand{\arraystretch}{0.8}
 \tabcolsep 0.5mm      
       \centering
       \caption{Pentaquark masses for $N_b$ = 1  for all possible configurations of singly bottom pentaquarks. Masses are in the units of MeV.}
       \begin{tabular}{ccccccccc}
\hline
\hline
 Quark Content &  S &  I &  Y & $SU(3)_f$ & $C_2(SU(3)_f)$ & $N_b$ & Our prediction(MeV) &  Reference \\
  \hline
  \\
  $P_{b0}^{++}$,  $P_{b0}^{+}$,  $P_{b0}^{0}$,  $P_{b0}^{-}$,  $P_{b0}^{--}$ & 1/2 & 2 & 4/3 & $[4]_{15}$ &  16/3 & 1 & 6650.62 $\pm$ 35.46 & 6780 \cite{Singly}\\
  
   & 3/2 & 2 & 4/3 & $[4]_{15}$ &  16/3 & 1 & 6719.62 $\pm$ 36.18 & 6746 \cite{Singly}\\
  \\
  \hline
  \\
  $P_{b1}^+$, $P_{b1}^0$, $P_{b1}^-$, $P_{b1}^{--}$ & 1/2 & 3/2 & 1/3 & $[4]_{15}$ &  16/3 & 1 & 6750.25 $\pm$ 35.46 & 6657 \cite{Singly} \\
  
  & 3/2 & 3/2 & 1/3  & $[4]_{15}$ &  16/3 & 1 & 6819.25 $\pm$ 35.74 & 6895 \cite{Singly} \\
 \\
  \hline
  \\
 $P_{b2}^{0}$,  $P_{b2}^{-}$,  $P_{b2}^{--}$ & 1/2 & 1 & -2/3 & $[4]_{15}$ &  16/3 & 1 & 6849.89 $\pm$ 35.23 & 6707 \cite{Singly} \\
  & 3/2 & 1 & -2/3 & $[4]_{15}$ &  16/3 & 1 & 6918.89 $\pm$ 35.50 & 6972 \cite{Singly}\\
  \\
  \hline
  \\
  $P_{b3}^{-}$,  $P_{b3}^{--}$ & 1/2 & 1/2 & -5/3 & $[4]_{15}$ &  16/3 & 1 & 6949.52 $\pm$ 35.20 & 6899 \cite{Singly}\\
  & 3/2 & 1/2 & -5/3 & $[4]_{15}$ &  16/3 & 1 & 7018.52 $\pm$ 35.47 & 7079 \cite{Singly}\\
  \\
  \hline
  \\
  $P_{b4}^{--}$ & 1/2 & 0 & -8/3 & $[4]_{15}$ &  16/3 & 1 & 7049.15 $\pm$ 35.38 & 7229 \cite{Singly}\\
  & 3/2 & 0 & -8/3 & $[4]_{15}$ &  16/3 & 1 & 7118.15 $\pm$ 35.65 & 7193 \cite{Singly}\\
 \hline
 \\
 $uuu\bar{u}b$, $uud\bar{u}b$, $udd\bar{u}b$ &   1/2 &  0 &  2/3 & $[4]_{15}$ & 16/3 & 1& 6574.82$\pm$ 35.14 \\
  $ddd\bar{u}b$, $uud\bar{d}b$  &   1/2 &  1 &  2/3 & $[4]_{15}$ & 16/3 & 1&  6638.82$\pm$ 35.23 \\
  $udd\bar{d}b$, $uuu\bar{d}b$  &   1/2 & 2 &  2/3 & $[4]_{15}$ & 16/3 & 1&  6766.82$\pm$ 35.97 \\
 $uus\bar{s}b$, $uds\bar{s}b$, $dds\bar{s}b$ &  3/2 &  0 &   2/3 & $[4]_{15}$ & 16/3 & 1  & 6643.82$\pm$ 35.42  \\
  &  3/2 &  1 &   2/3 & $[4]_{15}$ & 16/3 & 1  & 6707.82$\pm$ 35.50  \\
  &  3/2 &  2 &   2/3 & $[4]_{15}$ & 16/3 & 1  & 6835.82$\pm$ 36.23 
   \\
 \hline
 \\
 $uuu\bar{s}b$, $uud\bar{s}b$ &  1/2 &  1/2 &  5/3 & $[4]_{15}$ & 16/3 & 1  & 6421.85$\pm$ 35.20 \\
 $udd\bar{s}b$, $ddd\bar{s}b$ &  1/2 &  3/2 &  5/3 & $[4]_{15}$ & 16/3 & 1  & 6517.85$\pm$ 35.42\\
  &  1/2 &  5/2 &  5/3 & $[4]_{15}$ & 16/3 & 1  & 6677.85$\pm$ 36.72 \\
  &  3/2 &   1/2 &   5/3 & $[4]_{15}$ & 16/3 &  1  & 6490.85$\pm$ 35.47 \\
  &  3/2 &   3/2 &   5/3 & $[4]_{15}$ & 16/3 &  1  & 6586.85$\pm$ 35.69 \\
  &  3/2 &   5/2 &   5/3 & $[4]_{15}$ & 16/3 &  1  & 6746.85$\pm$ 36.99 \\
    \\
    \hline 
    \\
 $uus\bar{u}b$, $uss\bar{s}b$, $dss\bar{s}b$ &  1/2 &  1/2 &   -1/3 & $[4]_{15}$ & 16/3 & 1  & 6759.79$\pm$ 35.15 \\
  $dds\bar{u}b$, $dds\bar{d}b$ &  1/2 &  3/2 &   -1/3 & $[4]_{15}$ & 16/3 & 1  & 6855.79$\pm$ 35.46 \\
  $uus\bar{d}b$, $uds\bar{u}b$ &  1/2 &  5/2 &   -1/3 & $[4]_{15}$ & 16/3 & 1  & 7015.79$\pm$ 36.92 \\
  & 3/2 &  1/2 &   -1/3 & $[4]_{15}$ & 16/3 & 1  & 6829.79$\pm$ 35.42  \\
  &  3/2 &  3/2 &   -1/3 & $[4]_{15}$ & 16/3 & 1  & 6924.79$\pm$ 35.74  \\
  & 3/2 &  5/2 &   -1/3 & $[4]_{15}$ & 16/3 & 1  & 7084.79$\pm$ 37.18  \\
    \\
  \hline
  \\
 $uss\bar{u}b$, $uss\bar{d}b$  &  1/2  &  0 &  -4/3 & $[4]_{15}$ &  16/3 &  1  & 6880.75$\pm$ 35.18 \\
 $dss\bar{d}c$  &  1/2  &  1 &  -4/3 & $[4]_{15}$ &  16/3 &  1  & 6944.75$\pm$ 35.23 \\
  &   3/2  & 0 & -4/3 & $[4]_{15}$ &   16/3 & 1  & 6949.75$\pm$ 35.45\\
   &   3/2  & 1 &  -4/3 & $[4]_{15}$ &   16/3 & 1  & 7013.75$\pm$ 35.51\\
   \\
  \hline
  \\
 $sss\bar{u}c$, $sss\bar{d}c$ &   1/2  &  1/2 &  -7/3 & $[4]_{15}$ & 16/3 &   1  & 7033.72$\pm$ 35.27\\
 &   3/2  &  3/2 &   -7/3 & $[4]_{15}$ & 16/3 &  1  &  7102.72$\pm$ 35.54\\
  \\
  \hline
  \hline
 \end{tabular}
   \label{tab:nb}
\end{table*}

\subsection{Effective Mass Scheme}
The magnetic moment encoding of the pentaquark provides useful information about the distribution of charge and magnetization within hadrons, allowing for a better understanding of their geometric configurations. In this study, we calculated the effective mass of quarks (anti-quarks) by considering their interactions with nearby quarks via a single gluon exchange scheme. We calculated the masses and magnetic moments of the singly heavy pentaquark states using effective quark masses, which helped us explore their inner structure. The mass of pentaquarks can be written in two ways \cite{VERMA}:
\begin{align}
        M_P =& \sum_{i=1}^5 m_i^{eff}  \\
   M_P =& \sum_{i=1}^5 m_i + \sum_{i<j} b_{ij} s_i.s_j
   \label{Effective mass}
\end{align}
 here, $s_i$ and $s_j$ represent the spin operator for the $i^{th}$ and $j^{th}$ quarks (antiquark), and $m_i^{eff}$ represents the effective mass for each of the quark (antiquark) and $b_{ij}$ are the hyperfine interaction terms. For different quarks inside the pentaquark, effective masses equations are:
 \begin{equation}
 m_1^{eff} = m_1 + \alpha b_{12} + \beta b_{13} + \gamma b_{14} + \eta b_{15}
 \end{equation}
 
 \begin{equation}
  m_2^{eff} = m_2 + \alpha b_{12} + \beta^{'} b_{23} + \gamma^{'} b_{24} + \eta^{'} b_{25}
  \end{equation}

   \begin{equation}
   m_3^{eff} = m_3 + \beta b_{13} + \beta^{'} b_{23} + \gamma^{''} b_{34} + \eta^{''} b_{35} 
   \end{equation}

   \begin{equation}
    m_4^{eff} = m_4 + \gamma b_{14} + \gamma^{'} b_{24} + \gamma^{''} b_{34} + \eta^{'''} b_{45} 
 \end{equation}

  \begin{equation}
    m_5^{eff} = m_5 + \eta b_{15} + \eta^{'} b_{24} + \eta^{''} b_{34} + \eta^{'''} b_{45} 
 \end{equation}
  Here, 1, 2, 3, 4, and 5 stand for $u$, $d$, $s$, $c$, and $b$ quarks. These equations get modified if we consider two/three/four/five identical quarks. Therefore, products of spin quantum number are defined as \cite{Rohit}:
\begin{align*}
      s_{i}\cdot s_{j}=&  +1/4 \rightarrow \hspace{0.3cm}\uparrow \uparrow\\
       =& -1/2 \rightarrow \hspace{0.3cm} \uparrow \downarrow\\
       =& -1/4 \rightarrow \hspace{0.3cm}\downarrow \downarrow
\end{align*}
Therefore, for spin-3/2 ($\uparrow\uparrow\uparrow\uparrow\downarrow$) pentaquarks,
\begin{equation}
s_1 \cdot s_2 = s_1 \cdot s_3 =  s_1 \cdot s_4 = 1/4 
\end{equation}

\begin{equation}
s_2 \cdot s_3 = s_2 \cdot s_4 =  s_3 \cdot s_4 =  1/4 
\end{equation}

\begin{equation}
s_1 \cdot s_5 = s_2 \cdot s_5 = s_3 \cdot s_5 = s_4 \cdot s_5 =  = -1/2 
\end{equation}
and parameters are calculated by using eq.\eqref{Effective mass}:
 \begin{equation}
     \alpha = \beta = \gamma = 1/8
 \end{equation}

\begin{equation}
\beta^{'} = \gamma^{'} = \gamma^{''}= 1/8
\end{equation}

\begin{equation}
\eta = \eta^{'} = \eta^{''} = \eta^{'''} =  -1/4
\end{equation}
Similarly, for the case of spin-1/2 ($\uparrow\uparrow\uparrow\downarrow\downarrow$) pentaquarks,
\begin{equation}
s_1 \cdot s_2 = s_1 \cdot s_3 =  s_2 \cdot s_3 = 1/4 
\end{equation}

\begin{equation}
s_1 \cdot s_4 = s_1 \cdot s_5 =  s_2 \cdot s_4 =s_2 \cdot s_5= s_3 \cdot s_5 =  -1/2 
\end{equation}

\begin{equation}
s_4 \cdot s_5 = -1/4
\end{equation}
and parameters are calculated by using eq.\eqref{Effective mass}:
 \begin{equation}
     \alpha = \beta = \gamma = 1/8
 \end{equation}

\begin{equation}
\beta^{'} = \gamma^{'} = \gamma^{''} = \eta = \eta^{'} = \eta^{''} = -1/4
\end{equation}

\begin{equation}
\eta^{'''} =  -1/8
\end{equation}

Therefore, a modified form of the equation of effective mass for $J^P = 3/2^-$ ($\uparrow\uparrow\uparrow\uparrow\downarrow$) pentaquarks is defined as:
\begin{align}
M_{P_{{3/2}^-}} = m_1 + m_2 + m_3 + m_4 + m_5 + \frac{b_{12}}{4}  + \frac{b_{13}}{4}  + \frac{b_{14}}{4} \nonumber \\
 - \frac{b_{15}}{2}  + \frac{b_{23}}{4}  + \frac{b_{24}}{4}  - \frac{b_{25}}{2}  + \frac{b_{34}}{4}  - \frac{b_{35}}{2}  - \frac{b_{45}}{2}
 \label{eff2}
\end{align}
Similarly, we can write this for $J^P = 1/2^-$ ($\uparrow\uparrow\uparrow\downarrow\downarrow$) pentaquarks as:
\begin{align}
M_{P_{{1/2}^-}} = m_1 + m_2 + m_3 + m_4 + m_5 + \frac{b_{12}}{4}  + \frac{b_{13}}{4}  - \frac{b_{14}}{2} \nonumber \\
 - \frac{b_{15}}{2}  + \frac{b_{23}}{4}  - \frac{b_{24}}{2}  - \frac{b_{25}}{2}  - \frac{b_{34}}{2}  - \frac{b_{35}}{2}  - \frac{b_{45}}{4}
 \label{eff3}
\end{align}

The value of the quark masses are taken from Ref. \cite{Rohit}, and hyperfine interaction terms $b_{ij}$ are calculated using the effective mass equations \eqref{eff2}, and \eqref{eff3} by taking the masses from Gursey-Radicati mass formula as a reference and written as:
\begin{align}
     m_u = m_d = \hspace{0.3cm} 362 MeV, \hspace{0.3cm} m_s = \hspace{0.3cm} 539 MeV \nonumber \\
    m_c = \hspace{0.3cm} 1710 MeV, \hspace{0.3cm} m_b = \hspace{0.3cm} 5043 MeV
\end{align}
\begin{equation}
     b_{uu} = \hspace{0.3cm} b_{ud} = \hspace{0.3cm} b_{dd} = 115.11 MeV
 \end{equation}

 \begin{equation}
     b_{us} = \hspace{0.3cm} b_{ds} \hspace{0.3cm} = \left(\frac{m_u}{m_s}\right)b_{uu} = 77.30 MeV
 \end{equation}

\begin{equation}
     b_{uc} = \hspace{0.3cm} b_{dc} = \hspace{0.3cm} \left(\frac{m_u}{m_c}\right)b_{uu} = 24.36 MeV
 \end{equation}

\begin{equation}
     b_{ss} = \hspace{0.3cm} \left(\frac{m_u}{m_s}\right)^2 b_{uu} =  51.92 MeV
 \end{equation}

\begin{equation}
     b_{sc} = \hspace{0.3cm} \left(\frac{m_u^2}{m_s m_c}\right) b_{uu} = 16.36 MeV
\end{equation}

\begin{equation}
     b_{cc} = \hspace{0.3cm} \left(\frac{m_u}{ m_c}\right)^2 b_{uu} = 5.15 MeV
 \end{equation}

\begin{equation}
     b_{ub} = \hspace{0.3cm} b_{db} = \hspace{0.3cm} \left(\frac{m_u}{m_b}\right)b_{uu}  = 8.26 MeV
 \end{equation}

\begin{equation}
     b_{bb} = \hspace{0.3cm} \left(\frac{m_u}{m_b}\right)^2 b_{uu} =  0.59 MeV
 \end{equation}

 \begin{equation}
     b_{sb} = \hspace{0.3cm} \left(\frac{m_u}{m_s}\right) b_{ub} =  5.54 MeV
 \end{equation}

\begin{table*}\renewcommand{\arraystretch}{0.8}
 \tabcolsep 0.5mm
\centering
\caption{Table for the effective quark masses for singly heavy pentaquarks with different possible $J^P$ values. Here, $s$ denotes the number of strange quarks in the system, and  $m_u^*$, $m_d^*$, $m_s^*$ $m_c^*$, $m_{\Bar{c}}^*$ and $m_b^*$ are the effective quark masses for up, down, strange, charm, anti-charm and bottom quarks respectively. All the effective masses are in the units of MeV.}
\begin{tabular}{cccccccc}
       \hline
       \hline
         & \hspace{0.3cm} $m_u^*$/$m_d^*$ & \hspace{0.3cm}$m_s^*$ & \hspace{0.3cm}$m_c^*$ & \hspace{0.3cm}$m_{\Bar{c}}^*$ & \hspace{0.3cm}$m_b^*$&\hspace{0.3cm}$m_{\Bar{b}}^*$ & Total Effective Mass\\
         \hline
     \underline{Singly Charm Pentaquarks with $J^P = 1/2^-$} &  &   &   & \\
      $s=0$ & \hspace{0.3cm} 355.00  & \hspace{0.3cm} - & \hspace{0.3cm} 1688.69 & \hspace{0.3cm} 1688.69 & \hspace{0.3cm} - & -& 3029.04\\
      $s=1$ & \hspace{0.3cm} 365.36  &  \hspace{0.3cm} 478.98  & \hspace{0.3cm} 1689.69 & \hspace{0.3cm} 1689.69 & \hspace{0.3cm} - & - & 3264.75 \\
      $s=2$ & \hspace{0.3cm} 360.63  & \hspace{0.3cm} 541.25  & \hspace{0.3cm} 1691.69 & \hspace{0.3cm} 1691.69 & \hspace{0.3cm} - & - & 3439.54\\
      $s=3$ & \hspace{0.3cm} 355.91 & \hspace{0.3cm} 538.03 & \hspace{0.3cm} 1693.69 & \hspace{0.3cm} 1693.69 & \hspace{0.3cm} - & - & 3617.43\\
       $s=4$ & \hspace{0.3cm} - & \hspace{0.3cm} 534.91 & \hspace{0.3cm} 1695.69 & \hspace{0.3cm} 1695.69 & \hspace{0.3cm} - & - & 3798.43\\
        \hline
         \underline{Singly Charm Pentaquarks with $J^P = 3/2^-$} &  &   &   & \\
     $s=0$ & \hspace{0.3cm} 399.07   & \hspace{0.3cm} - & \hspace{0.3cm} 1685.64 & \hspace{0.3cm} 1685.64 & \hspace{0.3cm} - &- & 3281.94\\
      $s=1$ & \hspace{0.3cm} 394.39  &  \hspace{0.3cm} 563.89 & \hspace{0.3cm} 1687.64 & \hspace{0.3cm} 1687.64 & \hspace{0.3cm} - &- & 3434.59 \\
      $s=2$ & \hspace{0.3cm} 389.62  & \hspace{0.3cm} 560.72 & \hspace{0.3cm} 1689.64 & \hspace{0.3cm} 1689.64 & \hspace{0.3cm} - &- & 3590.34 \\
      $s=3$ & \hspace{0.3cm} 384.98  & \hspace{0.3cm} 557.53  & \hspace{0.3cm} 1691.64 & \hspace{0.3cm} 1691.64 & \hspace{0.3cm} - &- & 3749.20\\
    $s=4$ & \hspace{0.3cm} - & \hspace{0.3cm} 554.38 & \hspace{0.3cm} 1693.64 & \hspace{0.3cm} 1693.94 & \hspace{0.3cm} - & - & 3911.16\\
      \hline
       \underline{Singly Bottom Pentaquarks with $J^P = 1/2^-$} &  &   &   & \\
      $s=0$ & \hspace{0.3cm} 359.93  & \hspace{0.3cm} - & \hspace{0.3cm}- & \hspace{0.3cm}-&  \hspace{0.3cm} 5035.77 & \hspace{0.3cm} 5035.77 & 6390.21\\
      $s=1$  &  \hspace{0.3cm} 369.38 & \hspace{0.3cm} 480.43 & \hspace{0.3cm} - &\hspace{0.3cm} - &\hspace{0.3cm} 5036.11 & \hspace{0.3cm} 5036.11 & 6624.61 \\
      $s=2$ & \hspace{0.3cm} 364.66  & \hspace{0.3cm} 543.96 & \hspace{0.3cm} - & \hspace{0.3cm} - &\hspace{0.3cm} 5036.79 & \hspace{0.3cm} 5036.79 & 6796.75\\
      $s=3$ & \hspace{0.3cm} 359.93 & \hspace{0.3cm} 540.78  & \hspace{0.3cm} - & \hspace{0.3cm} - &\hspace{0.3cm} 5037.47 & \hspace{0.3cm} 5037.47 & 6972.01\\
    $s=4$ & \hspace{0.3cm} - & \hspace{0.3cm} 537.61 & \hspace{0.3cm} - & \hspace{0.3cm} - & \hspace{0.3cm} 5038.15 & \hspace{0.3cm} 5038.15 & 7150.37\\
      \hline
      \hline
       \underline{Singly Bottom Pentaquarks with $J^P = 3/2^-$} &  &   &   & \\
      $s=0$ & \hspace{0.3cm} 403.10 & \hspace{0.3cm} -  & \hspace{0.3cm} - & \hspace{0.3cm} - & \hspace{0.3cm} 5034.74 & \hspace{0.3cm} 5034.74 & 6647.15\\
      $s=1$ & \hspace{0.3cm} 398.37 &  \hspace{0.3cm} 566.60 & \hspace{0.3cm} - & \hspace{0.3cm} - & \hspace{0.3cm} 5035.42 & \hspace{0.3cm} 5035.42 & 6797.15 \\
      $s=2$ & \hspace{0.3cm} 393.64 & \hspace{0.3cm} 563.43  & \hspace{0.3cm} - & \hspace{0.3cm} - & \hspace{0.3cm} 5036.10 & \hspace{0.3cm} 5030.10 & 6950.26 \\
      $s=3$ & \hspace{0.3cm} 388.92 & \hspace{0.3cm} 560.25 & \hspace{0.3cm} - & \hspace{0.3cm} - & \hspace{0.3cm} 5036.78 &  \hspace{0.3cm} 5036.78 & 7106.47\\
       $s=4$ & \hspace{0.3cm} - & \hspace{0.3cm} 557.08 & \hspace{0.3cm} - & \hspace{0.3cm} - & \hspace{0.3cm} 5037.46 & \hspace{0.3cm} 5037.46 & 7265.80\\
      \hline
      \hline
       \end{tabular}
        \label{tab:em}
   \end{table*}

 Using the values of hyperfine interaction terms $b_{ij}$ and quark masses, effective quark masses for $J^P$ = $1/2^-$, $3/2^-$ for singly heavy pentaquarks are reported in Table \ref{tab:em}. In the next subsection, we introduced the screened charge scheme to calculate the magnetic moments of singly heavy pentaquark states.

\subsection{Screened Charge Scheme}
The charge of a quark within an exotic hadron may be influenced in the same way that its mass is affected by its surroundings. This can be understood by taking an example: when a soft photon probes a quark inside a pentaquark, its charge may be screened due to its existence of neighboring quarks. This effect is analogous to the surrounding electron cloud protecting the nuclear charge of a helium atom. We assume the effective charge is directly related to the shielding quark charge. Therefore, the effective charge of the quark 'a' in pentaquark (a,f, x, y, z) can be written as: 
\begin{equation}
    e_a^P = e_a + \alpha_{af} e_f + \alpha_{ax} e_x + \alpha_{ay} e_y + \alpha_{az} e_z
    \label{screen1}
    \end{equation}
In a similar manner, 
\begin{equation}
    e_f^P = e_f + \alpha_{fa} e_a + \alpha_{fx} e_x + \alpha_{fy} e_y + \alpha_{fz} e_z,
    \label{screen2}
    \end{equation}

    \begin{equation}
    e_x^P = e_x + \alpha_{xa} e_a + \alpha_{xf} e_f + \alpha_{xy} e_y + \alpha_{xz} e_z,
    \label{screen3}
    \end{equation}

\begin{equation}
    e_y^P = e_y + \alpha_{ya} e_a + \alpha_{yf} e_f + \alpha_{yx} e_x + \alpha_{yz} e_z,
    \label{screen4}
    \end{equation}

    \begin{equation}
    e_z^P = e_z + \alpha_{za} e_a + \alpha_{zf} e_f + \alpha_{zx} e_x + \alpha_{zy} e_y
    \label{screen5}
    \end{equation}
where $e_a$, $e_f$, $e_x$, $e_y$ and $e_z$ are the charges of the quarks. By considering the isospin symmetry, we get various relations:
\begin{align}
    \alpha_{af} = \alpha_{fa}, \hspace{0.3cm}\alpha_{ax} = \alpha_{xa}, \hspace{0.3cm} \alpha_{ay} = \alpha_{ya}, \hspace{0.3cm} \alpha_{az} = \alpha_{za}
\end{align}
Thus,
\begin{align}
    \alpha_{uu} = \alpha_{ud} = \alpha_{dd} = \alpha_1 \\ \nonumber
    \alpha_{us} = \alpha_{ds} = \beta_1 \\ \nonumber
    \alpha_{ss} = \beta_2
\end{align}
 In charm sector,
 \begin{align}
     \alpha_{uc} = \alpha_{dc} = \beta_3 \\ \nonumber
     \alpha_{sc} = \alpha_2, \hspace{0.3cm} \alpha_{cc} = \alpha_3
 \end{align}
 In a similar manner, for the bottom sector,
 \begin{align}
     \alpha_{ub} = \alpha_{db} = \beta_4, \hspace{0.3cm} \alpha_{sb} = \alpha_4 \\ \nonumber
     \alpha_{cb} = \beta_5, \hspace{0.3cm} \alpha_{bb} = \alpha_5
 \end{align}
If we use SU(3) symmetry, then these parameters can be reduced further as:
\begin{equation}
    \alpha_1 = \beta_1 = \beta_2
\end{equation}
Now, for the calculation of screening parameter $\alpha_{ij}$, we used Ansatz formalism, 
\begin{equation}
    \alpha_{ij} = \mid{\frac{m_i - m_j}{m_i + m_j}}\mid \times \delta
\end{equation}
where, $m_i$ and $m_j$ are the respective quark masses ($i, j = u, d, s, c, b$) and $\delta$ = 0.81 \cite{Bains}. The values of parameters help us to predict the magnetic moments of singly heavy pentaquark states. By putting these values of parameters in effective charge equations \eqref{screen1}, \eqref{screen2}, \eqref{screen3}, \eqref{screen4}, and \eqref{screen5} and by introducing the magnetic moment operator:
\begin{equation}
    \mu = \sum_i \frac{e_i^P}{2 m_i^{eff}} \sigma_i
\end{equation}
Now, the magnetic moment of the multiquark system consists of two parts:
 \begin{equation}
     \Vec{\mu} = \Vec{\mu}_{spin} + \Vec{\mu}_{orbit}
 \end{equation}
which can be written as:
\begin{equation}
    \Vec{\mu} = \hspace{0.3cm} \sum_i \mu_i (2 \Vec{s_i} + \Vec{l_i}) = \hspace{0.3cm} \sum_i \mu_i(2\Vec{s_i}) = \hspace{0.3cm} \sum_i \mu_i(\Vec{\sigma_i})
\label{Magnetic moment}
\end{equation}
Due to the absence of orbital excitations, magnetic moment depends only on the spin part. Therefore, by calculating the expectation value of Eq.\eqref{Magnetic moment} using pentaquark spin-flavor wavefunction, magnetic moments can be obtained:
\begin{equation}
    \mu = \bra{\Psi_{sf}}\Vec{\mu}\ket{\Psi_{sf}}
    \label{magnetic123}
\end{equation}

The expressions for the magnetic moments of pentaquarks can be followed using the above relations and written in Table \ref{tab: expressions}. Using these expressions, we calculated the magnetic moments of singly heavy pentaquarks. Using the  $\chi_4$ symmetry, we calculated the magnetic moments of spin-3/2 pentaquarks, and a linear combination of $\chi_6$ and $\chi_9$ symmetries for the magnetic moments of spin-1/2 spin pentaquarks, using the effective mass and screened charge schemes for both charm and bottom pentaquarks respectively.  The values for the magnetic moments of singly charm and bottom pentaquarks are reported in Tables \ref{tab: magnetic moment} and \ref{tab: magnetic moments}, respectively, by using the effective mass and screened charge schemes. 

\begin{table}[t]
    \centering
      \caption{Expressions for the magnetic moments of pentaquarks for possible spin symmetries with $J^P$ = $1/2^-$, $3/2^-$, $5/2^-$ (in $\mu_N$) \cite{young}.}
    \begin{tabular}{cc}
    \hline
    \hline
       Spin Basis  & \hspace{0.3cm} Magnetic Moment ($\mu$) \\
       \hline
        $\chi_1^P$  & \hspace{0.3cm}  $\mu_1$ + $\mu_2$ + $\mu_3$ + $\mu_4$ + $\mu_5$ \\
        \\
        $\chi_2^P$ & \hspace{0.3cm} $\frac{9}{10}$ ($\mu_1$ + $\mu_2$ + $\mu_3$ + $\mu_4$) -$\frac{3}{5}$ $\mu_5$\\
        \\
        $\chi_3^P$ & \hspace{0.3cm} $\frac{5}{6}$ ($\mu_1$ + $\mu_2$ + $\mu_3$) -$\frac{1}{2}$ $\mu_4$ + $\mu_5$\\
        \\
       $\chi_4^P$ & \hspace{0.3cm} $\frac{2}{3}$ ($\mu_1$ + $\mu_2$) - $\frac{1}{3}$ $\mu_3$ + $\frac{5}{6}$ ($\mu_4$ + $\mu_5$)\\
       \\
       
       $\chi_5^P$ & \hspace{0.3cm} $\mu_3$ + $\mu_4$ + $\mu_5$ \\
       \\
        $\chi_6^P$ & \hspace{0.3cm} $\frac{5}{6}$ ($\mu_1$ + $\mu_2$ + $\mu_3$) -$\frac{1}{3}$ ($\mu_4$ + $\mu_5$) \\
        \\
        $\chi_7^P$ & \hspace{0.3cm} $\frac{4}{9}$ ($\mu_1$ + $\mu_2$) - $\frac{2}{9}$ $\mu_3$ + $\frac{2}{3}$$\mu_4$ -  $\frac{1}{3}$ $\mu_5$   \\
        \\
       $\chi_8^P$ & \hspace{0.3cm}  $\frac{2}{3}$ ($\mu_3$ + $\mu_4$) - $\frac{1}{3}$ $\mu_5$ \\
       \\
        $\chi_9^P$ & \hspace{0.3cm} $\mu_5$  \\
        \\
        $\chi_{10}^P$ & \hspace{0.3cm} $\mu_5$ \\
        
        \hline
        \hline
    \end{tabular}
  \label{tab: expressions}
\end{table}

\begin{table*}[]
    \centering
    \caption{Magnetic moments of singly charm pentaquark states using the effective mass, screened charge,
    and effective mass plus screened charge scheme together for 
 $J^P = 1/2^-, 3/2^-$, respectively. All values of magnetic moments are in the units of $\mu_N$.}
    \renewcommand{\arraystretch}{1.0}
 \tabcolsep 0.01cm
    \begin{tabular}{|c|cc|cc|cc|}
    \hline
   
      Quark Content  & \hspace{0.3cm}Effective Mass Scheme &  \multicolumn{2}{c}{\hspace{0.2cm} \hspace{0.4cm} Screened Charge Scheme} & \multicolumn{2}{c}{\hspace{0.2cm} \hspace{0.4cm} Effective mass + Screened Charge Scheme} &\\
      \hline
        & $1/2^-$  & $3/2^-$  & $1/2^-$ & $3/2^-$  & $1/2^-$ & $3/2^-$  \\
        \hline

$uuuu\Bar{c}$  & 3.38 & 2.56  & 2.04 & 1.83 & 1.99 & 1.70 \\
$uuud\Bar{c}$  & 4.53 & 0.60  & 2.71 & -0.56  & 2.94 & -0.50 \\
$uudd\Bar{c}$ & 2.33 & 1.38  & 0.36 & 0.06  & 0.55 & 0.04 \\
$uddd\Bar{c}$  & 0.13 & -0.17  & -1.99 & -1.90 & -1.83 & -1.71  \\
$dddd\Bar{c}$  & -2.06 & -1.74 & -4.34 & -3.85  & -4.23 & -3.58 \\
$uuus\Bar{c}$  & 4.25 & 0.81  & 1.96 & 0.36  & 1.97 & 0.33  \\
$uuds\Bar{c}$  & 2.11 & 1.60  & -0.29 & 0.75  & -0.25 & 0.65 \\
$udds\Bar{c}$  & -0.02 & -0.02  & -2.55 & -1.44  & -2.49 & -1.39 \\
$ddds\Bar{c}$  & -2.17 & -1.56  & -4.81 & -3.64  & -4.72 & -3.44 \\
$uuss\Bar{c}$  & 2.30 & 1.55  & 0.46 & 0.26  & 0.40 & 0.22 \\
$udss\Bar{c}$  & 0.13 & -0.05  & -2.02 & -1.84  & -2.12 & -1.75 \\
$ddss\Bar{c}$  & -2.03 & -1.65 & -4.51 & -3.95  & -4.65 & -3.74  \\
$usss\Bar{c}$  & 0.13 & 0.12 & -1.10 & -1.09  & -1.55 & -1.08 \\
$dsss\Bar{c}$  & -2.03 & -1.50 & -3.83 & -3.38 & -4.43 & -3.26 \\
$ssss\Bar{c}$  & -1.51 & -1.34 & -2.76 & -2.52  & -2.78 & -2.47 \\
 %%%%%%%%%%%%%%%%%%%%%%%%%%%%%%%%%
$uuu\Bar{u}c$  & 5.23 & 0.57 & -2.94 & 2.58  & 2.99 & 2.40 \\
$uud\Bar{u}c$  & 3.03 & 1.35 & 5.01 & 3.20 & 5.10 & 2.94 \\
$udd\Bar{u}c$  & 0.83 & -0.21 & 2.66 & 1.23  & 2.70 & 1.13 \\
$ddd\Bar{u}c$  & -1.36 & -1.78  & 0.30 & -0.73  & 0.31 & -0.67  \\
$uus\Bar{u}c$ & 3.46 & 1.28  & 4.88 & 2.90  & 5.10 & 2.70 \\
$uds\Bar{u}c$  & 1.23 & -0.29  & 2.30 & 1.03  & 2.45 & 0.96 \\
$dds\Bar{u}c$  & -0.99 & -1.88  & -0.28 & -0.84 & -0.20 & -0.77 \\
$uss\Bar{u}c$  & 1.48 & -0.14  & -2.32 & 0.80  & -2.49 & 0.76 \\
$dss\Bar{u}c$  & -0.77 & -1.75  & -0.49 & -1.25  & -0.42 & -1.17 \\
$sss\Bar{u}c$  & -0.50 & -1.60  & -0.32 & -1.23  & -0.23 & -1.17 \\
$uuu\Bar{d}c$  & 4.26 & 2.53  & 6.69 & 4.98  & -6.62 & 4.61 \\
$uud\Bar{d}c$  & 2.06 & 3.31 & 4.34 & 5.60 & 4.23 & 5.15 \\
$udd\Bar{d}c$  & -0.13 & 1.74  & 1.99 & 3.63 & 1.83 & 3.34 \\
$ddd\Bar{d}c$  & -2.33 & 0.17 & -0.36 & 1.66 & -0.55 & 1.52 \\
$uus\Bar{d}c$  & 2.36 & 3.27  & 4.44 & 4.92  & 4.42 & 4.52 \\
$uds\Bar{d}c$  & 0.13 & 1.68 & 1.85 & 3.34  & 1.77 & 3.12 \\
$dds\Bar{d}c$  & -2.01 & 0.09  & -0.28  & 1.46 & -0.30 & 1.36 \\
$uss\Bar{d}c$  & 1.88 & 1.86  & 2.11 & 3.30  & -2.08 & 3.10 \\
$dss\Bar{d}c$  & -0.37 & 0.25  & -0.70 & 1.24 & -0.83 & 1.16 \\
$sss\Bar{d}c$  & -0.22 & 0.42 & -0.61 & 1.32 & -0.76 & 1.26\\
$uuu\Bar{s}c$  & 4.13 & 2.35 & 6.63 & 4.97 & 6.71 & 4.70 \\
$uud\Bar{s}c$  & 2.08 & 3.15 & 4.15 & 5.37 & 4.23 & 5.03 \\
$udd\Bar{s}c$  & -0.13 & 1.56  & 1.89 & 3.17 & 1.85 & 2.98 \\
$ddd\Bar{s}c$  & -2.23 & -0.02 & -0.57 & 0.97 & -0.52 & 0.92 \\
$uus\Bar{s}c$  & 1.25 & 3.10 & 3.36 & 5.25 & 3.29 & 4.96 \\
$uds\Bar{s}c$  & 0.56 & 1.49 & 2.79 & 3.14 & 2.89 & 2.98 \\
$dds\Bar{s}c$  & 0.21 & -0.11 & 1.32 & 1.03 & 1.30 & 1.00 \\
$uss\Bar{s}c$  & -0.13 & 1.67 & 0.79 & 3.14 & 0.89 & 3.01 \\
$dss\Bar{s}c$  & -0.11 & 0.04  & -0.79 & 0.85 & 0.76 & 0.83 \\
$sss\Bar{s}c$  & -0.10 & 0.21 & 0.59 & 0.87 & 0.55 & 0.85 \\
\hline
    \end{tabular}
    \label{tab: magnetic moment}
\end{table*}

\begin{table*}[]
    \centering
     \caption{Magnetic moments of singly bottom pentaquark states using the effective mass, screened charge, and effective mass plus screened charge scheme together for $J^P = 1/2^-, 3/2^-$ respectively. All values of magnetic moments are in the units of $\mu_N$.}
    \renewcommand{\arraystretch}{1.0}
 \tabcolsep 0.01cm
    \begin{tabular}{|c|cc|cc|cc|}
    \hline
   
     Quark Content  & \hspace{0.3cm}Effective Mass Scheme &  \multicolumn{2}{c}{\hspace{0.2cm} \hspace{0.4cm} Screened Charge Scheme} & \multicolumn{2}{c}{\hspace{0.2cm} \hspace{0.4cm} Effective mass + Screened Charge Scheme} &\\
      \hline
        & $1/2^-$  & $3/2^-$  & $1/2^-$ & $3/2^-$  & $1/2^-$ & $3/2^-$  \\
        \hline
$uuuu\Bar{b}$  & 3.81 & 2.89 & 5.32 & 4.61 & 5.35 & 4.17 \\
$uuud\Bar{b}$  & 4.68 & 0.95  & 6.10  & 7.00 & 6.13 & 2.12 \\
$uudd\Bar{b}$  & 2.50 & 1.73  & 3.85 & 3.10  & 3.87 & 2.79  \\
$uddd\Bar{b}$  & 0.33 & 0.18  & 1.60 & 1.26  & 1.61 & 1.13  \\
$dddd\Bar{b}$  & -1.84 & -1.37  & -0.64 & -0.57  & -0.64 & -0.52  \\
$uuus\Bar{b}$  & 4.49 & 1.16 & 5.54 & 2.72 & 5.42 & 2.50 \\
$uuds\Bar{b}$  & 3.11 & 1.94 & 4.79 & 3.24 & 4.86 & 2.96 \\
$udds\Bar{b}$  & 1.75 & 0.37 & 2.83 & 1.17  & 2.96 & 1.06  \\
$ddds\Bar{b}$  & -1.96 & -1.19  & -0.52 & -0.88  & -0.50 & -0.83 \\
$uuss\Bar{b}$  & 2.16 & 1.89 & 3.63 & 2.94 & 3.79 & 2.72 \\
$udss\Bar{b}$  & 1.78 & 0.30 & 2.62 & 0.97 & 2.56 & 0.90 \\
$ddss\Bar{b}$  & -1.86 & -1.28 & -0.98 & -0.99 & -0.95 & -0.93 \\
$usss\Bar{b}$  & 0.85 & 0.47 & 2.38 & 1.28  & 2.35 & 1.19 \\
$dsss\Bar{b}$  & -1.87 & -1.13 & -0.91 & -0.87 & -0.92 & -0.83 \\
$ssss\Bar{b}$  & 1.68 & -0.97 & 2.23 & -0.45 & 2.16 & -0.44 \\
 %%%%%%%%%%%%%%%%%%%%%%%%%%%%%%%%%
$uuu\Bar{u}b$  & 4.77 & 0.20 & 6.10 & -0.72  & 6.16 & -0.63 \\
$uud\Bar{u}b$  & 2.59 & 0.98  & 3.85 & 0.03  & 3.90 & 0.02 \\
$udd\Bar{u}b$  & 0.42 & -0.56  & 1.60 & -1.80  & 1.64 & -1.63 \\
$ddd\Bar{u}b$  & -1.75 & -2.12 & -0.64 & -3.64  & -0.61 & -3.29 \\
$uus\Bar{u}b$ & 2.86 & 0.92 & 3.95 & -0.49 & 4.06 & -0.43  \\
$uds\Bar{u}b$  & 0.65 & -0.64 & 1.47 & -2.23 & 1.53 & -2.02 \\
$dds\Bar{u}b$  & -1.54 & -2.14 & -1.00 & -3.98  & -0.98 & -3.62 \\
$uss\Bar{u}b$  & 0.90 & -0.49 & 1.72 & -2.04  & 1.82 & -1.89 \\
$dss\Bar{u}b$  & -1.32 & -2.08 & -0.98 & -3.97  & -0.96 & -3.68 \\
$sss\Bar{u}b$  & -1.08 & -1.94 & -0.83 & -3.65 & -0.81 & -3.45 \\
$uuu\Bar{d}b$  & 3.92 & 2.14 & 3.33 & 1.54 & 3.32 & 1.40 \\
$uud\Bar{d}b$  & 1.75 & 2.92 & 1.17 & 2.30  & 1.15 & 2.07 \\
$udd\Bar{d}b$  & -0.42 & 1.37 & -1.07 & 0.46 & -1.10 & 0.41 \\
$ddd\Bar{d}b$  & -2.59 & -0.18 & -3.32 & -1.37 & -3.36 & -1.24 \\
$uus\Bar{d}b$  & 2.04 & 2.88 & 1.26 & 1.68 & 1.28 & 1.54 \\
$uds\Bar{d}b$  & -0.15 & 1.31 & -1.20 & -0.06 & -1.23 & -0.04 \\
$dds\Bar{d}b$  & -2.36 & -0.26 & -3.68  & -1.80  & -3.75 & -1.63 \\
$uss\Bar{d}b$  & 0.11 & 2.59 & -0.43 & 0.31 & -0.41 & 0.28 \\
$dss\Bar{d}b$  & -2.11 & -0.10 & -3.14 & -1.61  & -3.20 & -1.57 \\
$sss\Bar{d}b$  & -1.85 & 0.06 & -3.01 & -1.24  & -3.04 & -1.20 \\
$uuu\Bar{s}b$  & 3.97 & 1.98  & 3.09 & 2.09  & 3.00 & 1.94 \\
$uud\Bar{s}b$  & 1.86 & 2.76 & 0.74 & 2.61 & 0.90 & 2.40 \\
$udd\Bar{s}b$  & -0.25 & 1.19 & -1.21 & 0.55 & -1.20 & 0.50 \\
$ddd\Bar{s}b$  & -2.37 & -0.37 & -3.36 & -1.51 & -3.29 & -1.39 \\
$uus\Bar{s}b$  & 2.12 & 2.71 & 1.29 & 2.26 & 1.26 & 2.11 \\
$uds\Bar{s}b$  & -0.01 & 1.12 & -1.08 & 0.29 & -1.08 & 0.28 \\
$dds\Bar{s}b$  & -2.16 & -0.46 & -3.50 & -1.68 & -3.47 & -1.54 \\
$uss\Bar{s}b$  & 0.22 & 1.30 & -0.60 & 0.71  & -0.61 & 0.66 \\
$dss\Bar{s}b$  & -1.95 & -0.30 & -3.22 & -1.44 & 3.23 & -1.36 \\
$sss\Bar{s}b$  & -1.71 & -0.14  & -2.56 & -0.91 & -2.57 & -0.88 \\
\hline
    \end{tabular}
 \label{tab: magnetic moments}
\end{table*}

\section{Results and Discussion}
By taking into consideration the recent discoveries of singly heavy tetraquarks by LHCb collaboration having quark content $u\bar{d}c\bar{s}$ and $\bar{u}dc\bar{s}$ with the significance of 6.5 $\sigma$ and 8 $\sigma$ respectively \cite{2900}, we estimated the masses and magnetic moments of singly heavy pentaquark structures. Considering the SU(3) flavor representation, we classified the singly heavy pentaquarks into the $15^{'}$-plet and $15^{''}$ representation. Further, we studied the possible spin symmetries for pentaquark states using SU(2) spin representation and the Young tableau technique. There is one spin symmetry for spin-5/2 pentaquarks, four spin symmetries for spin-3/2 pentaquarks, and five spin symmetries for spin-1/2 pentaquarks. By considering the relevant symmetry for possible spin configurations, magnetic moments of singly heavy pentaquarks are calculated, and corresponding expressions for spin symmetries are written in Table \ref{tab: expressions}. Further, to find the masses of singly heavy pentaquarks, we used the extension of the Gursey-Radicati mass formula and the effective mass scheme. The extension of the Gursey- Radicati mass formula seems useful for studying the exotic hadronic states. Many works have been proposed using the extension of the Gursey-Radicati mass formula \cite{HOLMA, Santo, Sharma_2023}. Further, by using the effective mass scheme, we calculated the effective masses of each quark and the hyperfine interaction terms that describe the interaction of quarks with their neighboring quarks. We calculated the masses of pentaquark states using the effective quark masses and hyperfine interaction terms. effective mass scheme is a theoretical framework that proves valuable, especially when dealing with the magnetic moments of particles. The effective mass scheme is particularly relevant when particles behave as if they have a mass different from their rest mass due to their interaction with surrounding fields. Therefore, we studied the magnetic moments of the singly heavy pentaquark states using the effective mass and screened charge techniques. The mass spectrum of singly charm and bottom pentaquarks are reported in Tables \ref{tab:nc} and \ref{tab:nb} using the extension of the Gursey-Radicati mass formula, and corresponding comparisons are also reported. In Table \ref{tab:em}, singly heavy pentaquark masses are reported using the effective mass scheme. Our predicted masses are in reasonable agreement with the available data. For e.g., if we compare our predicted masses with the result obtained by H1 collaboration, they estimated the mass of $3099\pm 3 \pm 5$ MeV, whereas our estimation using the extension of the GR mass formula for respective quark content is 3185.64 $\pm$ 20.91 MeV and 3029.04 MeV using the effective mass scheme, which shows reasonable agreement with the results obtained by H1 collaboration. Further, Results of Ref. \cite{Singly} also shows good agreement with our predictions. To calculate the magnetic moments, we used the $\chi_4^P$ spin basis for spin-$3/2^-$ pentaquarks and $\chi_6^P$ and $\chi_9$ spin basis for spin-$1/2^-$ pentaquarks. The magnetic moment values for singly charm and bottom pentaquarks are reported in Tables \ref{tab: magnetic moment} and \ref{tab: magnetic moments}, respectively. Magnetic moments are calculated using the effective mass technique, screened charge technique, and both techniques together. We compared our results with the available theoretical data for singly heavy pentaquarks, showing reasonable agreement with our results. Our analysis of mass and magnetic moments may provide useful evidence for future experimental studies of these states.

\bibliography{main}% Produces the bibliography via BibTeX.

\end{document}